# Toward deep-learning-assisted spectrally-resolved imaging of magnetic noise

Fernando Meneses[1], David F. Wise[3,4], Daniela Pagliero[1], Pablo R. Zangara[5,6], Siddharth Dhomkar[3,†], and Carlos A. Meriles[1,2,†]

[1]*Department. of Physics, CUNY-City College of New York, New York, NY 10031, USA.*
[2]*CUNY-Graduate Center, New York, NY 10016, USA.*
[3]*London Centre for Nanotechnology, University College London, London WC1H 0AH, United Kingdom.*
[4]*Quantum Motion Technologies, Windsor House, Cornwall Road, Harrogate, England, HG1 2PW.*
[5]*Universidad Nacional de Córdoba, Facultad de Matemática, Astronomía, Física y Computación, Córdoba, Argentina.*
[6]*CONICET, Instituto de Física Enrique Gaviola (IFEG), Córdoba, Argentina.*

[†]*Corresponding authors. E-mail: s.dhomkar@ucl.ac.uk, cmeriles@ccny.cuny.edu.*

Recent progress in the application of color centers to nanoscale spin sensing makes the combined use of noise spectroscopy and scanning probe imaging an attractive route for the characterization of arbitrary material systems. Unfortunately, the traditional approach to characterizing the environmental magnetic field fluctuations from the measured probe signal typically requires the experimenter's input, thus complicating the implementation of automated imaging protocols based on spectrally resolved noise. Here, we probe the response of color centers in diamond in the presence of externally engineered random magnetic signals, and implement a deep neural network to methodically extract information on their associated spectral densities. Building on a long sequence of successive measurements under different types of stimuli, we show that our network manages to efficiently reconstruct the spectral density of the underlying fluctuating magnetic field with good fidelity under a broad set of conditions and with only a minimal measured data set, even in the presence of substantial experimental noise. These proof-of-principle results create opportunities for the application of machine-learning methods to color-center-based nanoscale sensing and imaging.

## I. INTRODUCTION

The ability to controllably engineer, manipulate, and probe near-surface color centers is driving a broad effort aimed at developing alternative nanoscale sensing and imaging platforms that leverage the color center susceptibility to changes in the local environment[1]. A paradigmatic example is the negatively-charged nitrogen-vacancy (NV) center in diamond, a paramagnetic point defect formed by a substitutional nitrogen impurity adjacent to a vacant lattice site[2]. Shallow NVs produced via low-energy ion implantation are now being routinely exploited as local probes to monitor magnetic[3], electric[4,5], and strain fields[6,7], as well as pH[8], temperature[9], and thermal conductivity[10]. Particularly relevant herein are applications to monitoring random magnetic or electric signals produced, for example, by small ensembles of fluctuating electron[11] or nuclear spins[12], or by thermally diffusing carriers[13]. While these measurements typically rely on the integrated color center response to the external random field, critical information on the local sample composition and dynamics can be derived from the noise spectral density $S(\omega)$ characterizing the environment[14,15]. Extracting $S(\omega)$ from the measured signal, however, is not straightforward as the process typically involves a delicate numerical deconvolution prone to error. This complication becomes a serious hurdle in situations where $S(\omega)$ must be repeatedly calculated, as in applications articulating noise spectroscopy and scanning imaging microscopy.

Recent work suggests this problem can be largely mitigated through the use of deep-learning algorithms explicitly adapted from advances in natural language processing, computer vision, and structured data analysis for applications in quantum physics and engineering[16,17]. Relevant examples include the use of deep neural networks (DNN) to improve quantum control operations[18-20], facilitate the processing of magnetic resonance data[21–25], and expose hidden patterns in scanning microscopy images[26]. In particular, a recent theoretical study[27] shows how DNNs can be trained to gather information on the random environment affecting a spin qubit, although the interplay between network performance, experimental uncertainty, and practical constraints (such as accessible noise bandwidth) remains unexplored.

Here, we monitor a diffraction-limited ensemble of NV center spins in diamond subjected to a random, time-dependent magnetic field purposely engineered to contain a pre-designed spectral density. We use a deep, feed-forward neural network previously trained to classify and/or extract $S(\omega)$ from the NV signal emerging upon application of a simple, two-pulse control protocol. Analyzing the response of the NV ensemble throughout a sequence of time-fluctuating fields of varying characteristics, we find the neural network can predict $S(\omega)$ with high accuracy over a broad spectral window, an ability we subsequently exploit to reconstruct the equivalent of a spectrally-resolved spatial map.

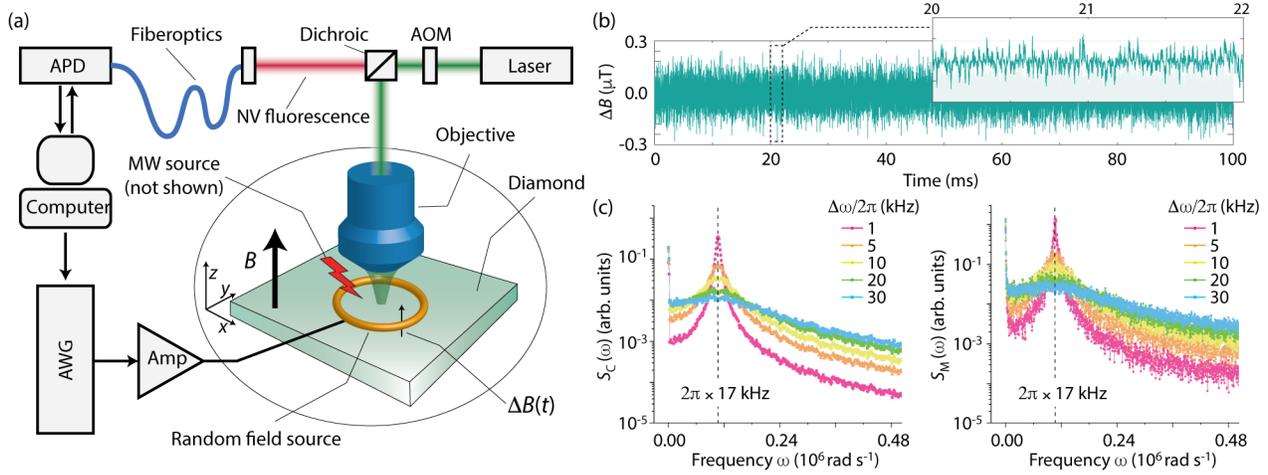

**Fig. 1. Synthesizing and probing magnetic noise.** (a) Schematics of the experimental setup. We use a [111] diamond with N and NV concentrations of ~1 ppm and ~20 ppb, respectively. NV$^-$ spin manipulation and detection are carried out via a home-made confocal microscope. A thin Cu wire overlaid on the diamond surface (not shown) serves as the MW source; experiments are carried out in the presence of a bias magnetic field **B** parallel to the [111] axis; we use a three-turn loop to produce a random magnetic field $\Delta \mathbf{B}(t)$ parallel to the bias field. (b) Example time trace of the generated random field $\Delta B(t)$; the upper insert is a zoomed view of the section in the dashed square. (c) Example colored noise with variable width; the left and right panels respectively reproduce the calculated and experimental spectral densities as derived from the current flowing through the loop. The light green curve corresponds to the conditions in (b). APD: Avalanche photo-detector; AOM: Acousto-optic modulator; AWG: Arbitrary wave-function generator.

## II. RESULTS
### Spin evolution in the presence of synthetic magnetic noise

In our experiments, we use optically-detected magnetic resonance (ODMR) to monitor the evolution of NV centers within a diffraction-limited volume in a [111], diamond with a 1 ppm nitrogen content (see Fig. 1a and Materials and Methods). To controllably introduce decoherence during coherent NV spin evolution, we first build on a source auto-spectral density to numerically synthesize a random time series[28], which we subsequently feed into an arbitrary wave-function generator and a home-made low-frequency amplifier to drive a three-turn, 1 mm diameter loop adjacent to the diamond crystal.

For illustration purposes, Fig. 1b shows an example trace of the time-dependent field $\Delta B(t)$ — effectively parallel to the bias field $B$ in the region of interest — as determined from the current circulating through the loop. We engineer $\Delta B(t)$ so as to capture the spectral density of choice: Fig. 1c displays representative examples for the case of a colored noise of variable width as calculated numerically from the input time series, or derived experimentally from the current in the loop (respectively, $S_C(\omega)$ and $S_M(\omega)$ shown in the left and right panels).

Applications of color centers to magnetometry typically rely on protocols where the spin probe is subjected to $n$ repetitions of a cycle comprising an inversion pulse sandwiched by free-evolution periods of duration $\tau/2$. Using $C(t)$ to denote the in-phase probe spin coherence amplitude after a total time $t = n\tau$, the noise spectrum relates to the "coherence functional" $\chi(t)$ via the integral equation[27,29]

$$\chi(t) = -\ln C(t) = \frac{1}{\pi}\int_0^\infty d\omega\, S(\omega)\, \frac{F_n(\omega t)}{\omega^2}. \quad (1)$$

In the above expression, $F_n(\omega t)$ is a filter function intrinsic to the *n*-pulse protocol, namely,

$$F_n(\omega t) = \left| 1 + (-1)^{n+1} e^{i\omega t} + 2\sum_{k=1}^{n}(-1)^k e^{i\omega t_k} \cos(\omega \tau_\pi/2) \right|^2, \quad (2)$$

where $t_k$ is the time of the *k*-th pulse in the train, and $\tau_\pi$ is the pulse duration. Importantly, the above formulas are based on the assumption of a Gaussian, stationary noise that leads to pure dephasing, all of which is consistent with our (classical) approach to introducing noise in the spin response.

Since numerically solving for $S(\omega)$ in Eq. (1) is notoriously difficult, alternative methods have been developed. The simplest approach[30,31] describes $F_n(\omega t)$ as a Dirac $\delta$ function centered at a frequency $\omega_0 = \pi/\tau$. Valid only for $n \gg 1$, this approximation fails when the inter-pulse separation and pulse duration become comparable (a limit where the filter function suffers from spectral delocalization and harmonic contributions difficult to account for). The result is a distorted spectral density that overemphasizes contributions at lower frequencies[27]. More accurate strategies have been developed[15], but they typically require collecting a much larger data set that includes the system time response under pulse trains with different number of pulses.

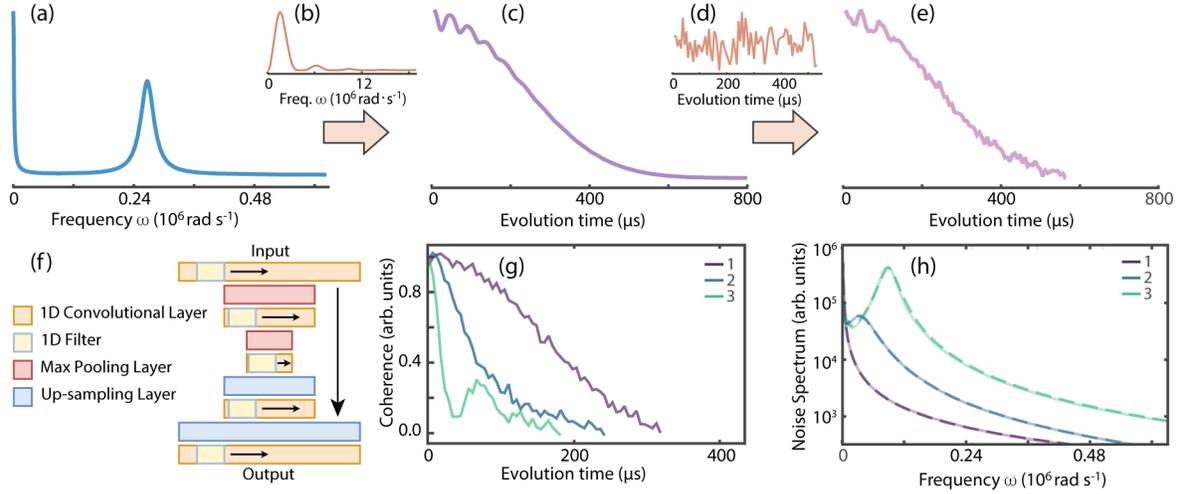

**Fig. 2. Deep neural network architecture and training procedure.** The top row illustrates the steps involved in training data generation. Simulated Hahn echo curves and the underlying noise spectra form the input-output pairs necessary for network training, specifically (a) an example noise spectrum, with an underlying $1/\omega$ form and a Lorentzian peak, characteristic of colored noise; (b) a filter function, determined by a Hahn echo pulse sequence with a given pulse spacing; (c) a decoherence curve, given by convolving the filter function at each time value with the noise spectrum in (a); (d) measurement noise sampled from a Gaussian distribution that mimics experimental error; (e) DNN input obtained by applying the measurement noise to the ideal decoherence curve. (f) A cartoon of the 1D convolutional auto-encoder architecture. Convolutions are filters applied to the input data (a coherence decay curve), the result of which is then down-sampled using a max pooling layer to select the largest of each neighboring pair of data points. The network down-samples the data to a bottleneck before up-sampling it back to meet the target data (a noise spectrum) dimension. (g) Three representative input coherence decay curves along with the derived noise spectra shown in (h); dashed lines show the network's predictions whilst solid lines show the true noise spectra.

Interestingly, Eq. (1) is remarkably asymmetric in the sense that although extracting $S(\omega)$ from $C(t)$ is involved, the converse operation — i.e., deriving the coherent response of the probe spin under the combined action of a known fluctuating field and control protocol — is straightforward. The latter suggests an artificial-intelligence-based approach designed to exploit the advantages of neural networks as universal function approximators[32].

**Artificial intelligence as a tool for noise spectroscopy**

Here, we implement a convolutional auto-encoder architecture (see Fig. 2 and Materials and Methods). We have found this approach provides similar or better performance than that observed previously[27] with significantly reduced parameter numbers and thus faster training time, likely a consequence of the auto-encoder capacity for greater parallelism[33]. The convolutional auto-encoder is a natural choice for the problem of noise spectrum extraction from coherence decay: By compressing the input data (in this case a coherence decay curve), the network learns to mine the essential information. The latter is, in turn, supplied to the decoder, which learns how to create the target output (in this case, the noise spectrum). We use the Adam optimizer to train the network with a variable learning rate gradually growing to reach a peak, before being reduced to zero as training progresses[34]. This approach was found to yield consistent training results whilst considerably reducing training time.

We implement our deep learning approach to noise spectroscopy via an optically detected Hahn echo sequence (i.e., the simplest magnetometry protocol in the $n$-pulse family highlighted above, Fig. 3a). To rationalize this choice, we first note that spin decoherence in color centers is largely dominated by the surrounding nuclear spin environment, which is uninteresting to most magnetometry applications and hence typically removed through echo formation. Improved suppression of the intrinsic nuclear field environment can be attained with longer pulse trains (i.e., $n > 1$), but multi-pulse protocols inherently feature narrower filter functions, implying that a more extensive data set — combining sequences with different inter-pulse separations *and* number of pulses — are needed to attain an accurate spectral density reconstruction. Free from harmonics and featuring a comparatively broader spectral filter function, the Hahn echo sequence we use herein appears, therefore, as a reasonable tradeoff, allowing us to drastically reduce the impact of the carbon bath on the NV response, while keeping the measurement protocol at its simplest.

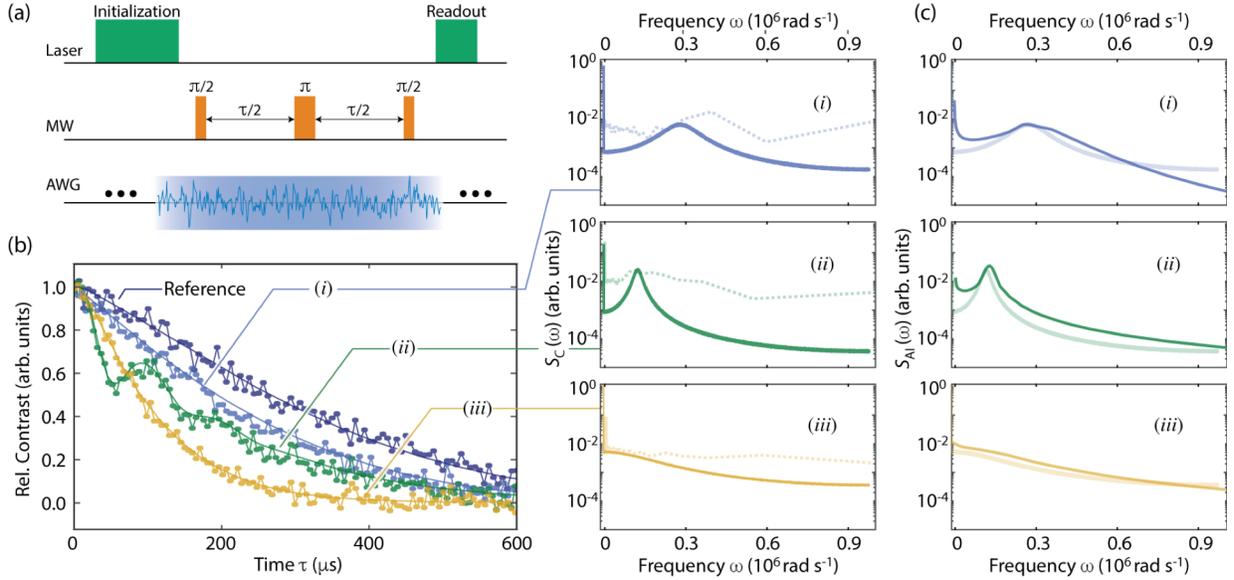

**Fig. 3. Artificial-intelligence-assisted noise spectroscopy.** (a) Schematics of the pulse sequence. We implement an optically-detected Hahn-echo protocol in the presence of a synthetic, random magnetic field. (b) Example NV spin echo curves in the presence of different random magnetic fields. The right-hand insert panels show the accompanying noise spectral densities $S_C(\omega)$; the dashed, faint traces indicate the spectral densities as calculated by taking the $\delta$-function approximation for the Hahn-echo filter function. Solid lines in the main plot indicate the system's response as calculated from the AI-derived spectral densities. All experiments are carried out in the presence of bias field $B = 36$ mT along the z-axis; the 'reference' trace in the main plot indicates the NV response for $\Delta B(t) = 0$. (c) AI-derived noise spectral densities $S_{AI}(\omega)$ for each of the Hahn-echo traces in (b); the faint lines reproduce $S_C(\omega)$ as presented in the right inserts of (b) for direct comparison.

In our experiments, we measure the NV coherent response as a function of the total Hahn-echo spin evolution time $\tau$. The dwell time between successive points in the trace (5.16 µs) is chosen to match the inverse of the bare $^{13}$C Larmor frequency in the bias magnetic field. This choice ensures detection of the NV coherence at the "revival" crests where the echo amplitude is maximum[35], and hence avoids modulations in the signal envelope that can otherwise obscure the effect of the noise field.

As an illustration, the main plot in Fig. 3b shows some example NV Hahn-echo signals in the presence of a time-dependent field $\Delta B(t)$ of variable root-mean-square amplitude, correlation time, and dominant frequency (see corresponding spectral densities on the side inserts); the observed response in the absence of a noise field — the 'reference' signal — is also included for comparison. Application of the time-dependent field shortens the NV spin coherence lifetime, which confirms $\Delta B(t)$ as the dominant source of NV spin dephasing.

Building on the results in Fig. 2, we now feed these echo time traces to our trained neural network and extract in each case the predicted noise spectral densities $S_{AI}(\omega)$. Fig. 3c displays the results: Comparison with the calculated densities $S_C(\omega)$ (side panels in Fig. 3b) shows reasonable agreement, even in cases such as (i) and (iii), where the NV response exhibits similar monotonic decay. Further, we find a significant improvement relative to reconstructions of the spectral densities based on a $\delta$-function approximation of the Hahn-echo filter function (faint dashed lines in the inserts of Fig. 3b). This advantage is comparable to that attained via high-accuracy multi-pulse-based protocols that take into account the presence of higher harmonics in the corresponding filter functions[15], though without the associated experimental overhead[27].

In a way, the accuracy of the AI-derived spectral densities is surprising because both training and validation of the network are based entirely on numerically computed data, hence suggesting that the uncertainty stemming from measurement error does not have an immediate impact on the network's ability to identify the correct answer. To better understand this observation, we systematically examine the effect of experimental noise in Fig. 4a, where we record the NV coherence traces for a variable number of times $N$ under a colored noise of constant central frequency (25 kHz), width (10 kHz), and root-mean-square amplitude (0.2 µT). Fig. 4b shows the AI-derived spectral densities, $S_{AI}(\omega)$, along with the numerically calculated density, $S_C(\omega)$, included here for reference. We find the network quickly captures the main spectral features and converges to a shape that reasonably reproduces that of the input magnetic noise. Further, as Figs. 4c through 4f show, the network output is stable once the experimental noise falls below a minimum threshold (of order ~15% in the present case). We hypothesize this 'robustness' to experimental noise may be a consequence of the decoupling between uncertainty in a given measurement and the shape of the overall NV response. In other words, the network continues to function correctly so long as the experimental noise does not obscure key, non-local features in the measured signal, a situation reminiscent of that of an 'expert' extracting information from a plot despite the dispersion in the data set.

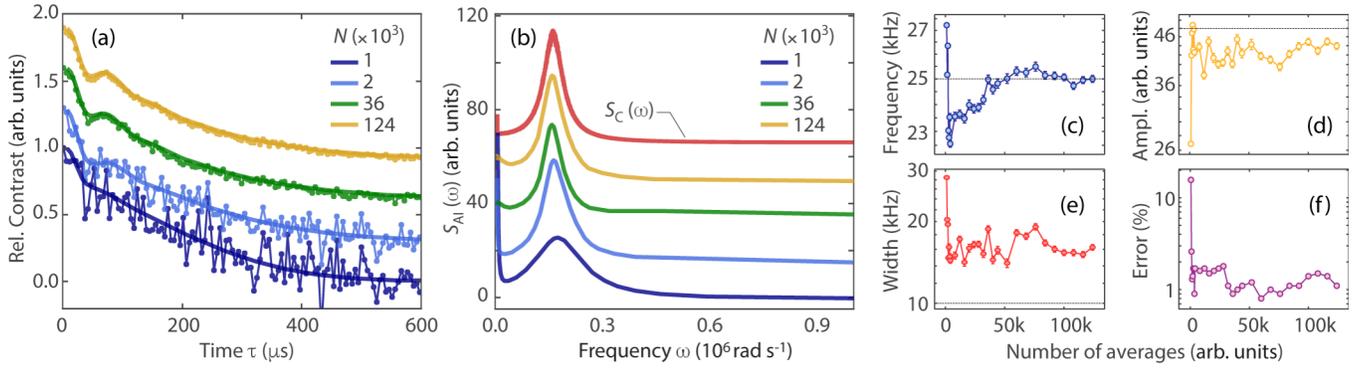

**Fig. 4. Dependence on experimental noise.** (a) NV Hahn-echo traces in the presence of colored magnetic noise (centered at $2\pi \times 25$ kHz, $2\pi \times 10$ kHz linewidth) after a variable number of averages $N$ (upper right corner). Solid lines indicate the AI-calculated coherences after AI-derivation of the spectral densities. (b) AI-derived spectral density for the experimental data in (a); the spectral density calculated from the input noise, $S_C(\omega)$, is also included for comparison. In (a) and (b), all traces have been displaced vertically for clarity. (c–e) Respectively, central frequency, amplitude, and linewidth of the AI-derived spectral density as a function of the number of experimental averages. Note all parameters quickly converge to values very much on par with those expected (dashed lines). (f) Percent error between the best experimental curve ($124 \times 10^3$ averages), taken as the reference, and the AI-predicted Hahn Echo curves using experimental curves with different number of averages. A single error point measures how far is the predicted curve from the reference and it is calculated by analyzing the squared differences between the two curves. The dashed lines in (c) through (e) indicate the expected values.

We formalize this notion through a principal component analysis[36] (PCA) over a large group of input coherence traces under the action of variable colored noise and/or "pink" noise (inversely proportional to the frequency). Rather than recording the output of the network, we instead examine the network's 'latent space', taking the output of the encoder before up-sampling. Although reduced in dimension in comparison to the input data, the latent space is still relatively high-dimensional. To better interpret it, we project the data onto the two dimensions along which it varies most, meaning that each input curve can be characterized by 2 parameters. Fig. 5a shows clustering of the data in two discernably distinct groups associated with monotonic and colored noise. Moreover, closer inspection of latent dimension 2 validates the network's ability to accurately distinguish between these types of noise, even in this reduced parameter space (Fig. 5b).

## Towards AI-assisted spectral imaging

Building on a better appreciation of the network capabilities and limitations, we now turn our attention to extracting $S(\omega)$ in a large data set comprising multiple NV Hahn-echo curves. Fig. 6a introduces our protocol: We measure the NV response $C^{(i)}(\tau)$ under a synthetic field $\Delta B^{(i)}(t)$, whose properties (i.e., root-mean-square amplitude, correlation time, etc.) we subsequently change from a starting set to the next one in a series $i = 1 \ldots N^2$. To more easily assess the network performance, we deliberately program the series $\{\Delta B^{(i)}\}$ so as to yield a pre-ordained "image" as we display the property of choice in $\{S_{\text{AI}}^{(i)}\}$ through a two-dimensional (2D) $N \times N$ array.

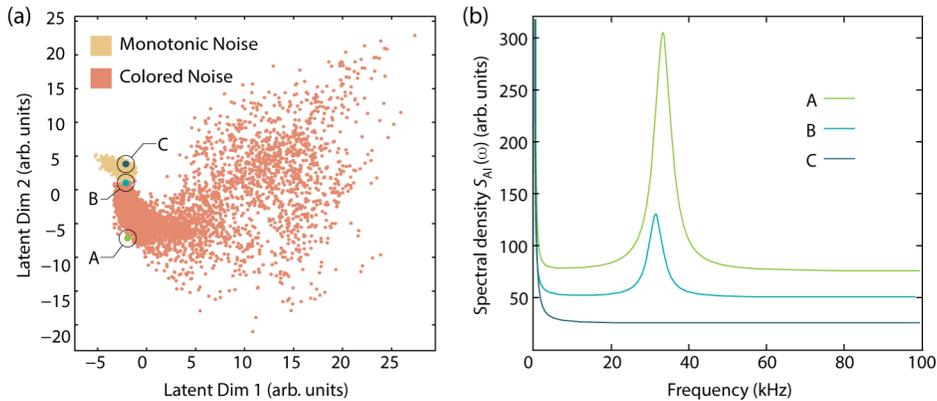

**Fig. 5. Principal component analysis of spin coherence traces.** (a) The test dataset in the latent space, projected onto two dimensions using PCA. The data points are colored according to their noise type, showing the clear distinction in the latent space for a trained network. (b) Example curves at different points in the latent space, showing the transition from monotonic to colored noise (points A, B, C in (a)).

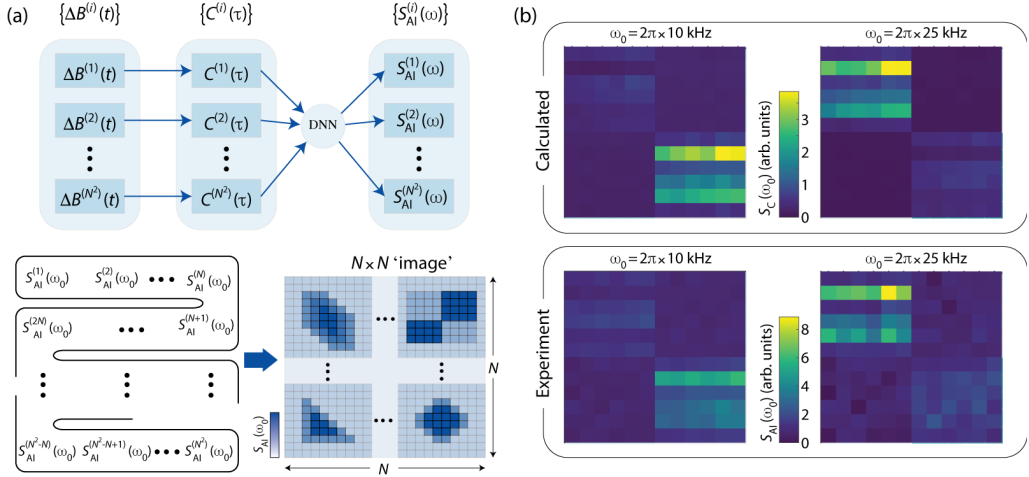

**Fig. 6. AI-assisted spectroscopic imaging: A proof of principle.** (a) (Top) We collect the NV Hahn-echo response $C^{(i)}(\tau)$ for fluctuating fields $\Delta B^{(i)}(t)$ of variable characteristics to form a data set with $i = 1 \ldots N^2$ components; in each case we resort to our DNN to derive the corresponding spectral densities $S_{\text{AI}}^{(i)}(\omega)$. (Bottom) We preordain the data series so as to introduce correlations in relevant spectral properties when presented in 2D $N \times N$ graph; in this illustration, we 'image' the spectral density $S_{\text{AI}}(\omega_0)$ at a given frequency $\omega_0$ to uncover local structure. (b) Predicted and AI-derived spectral densities (respectively, top and bottom panels) at two different frequencies $\omega_0 = 2\pi \times 10$ and $2\pi \times 25$ kHz (respectively, left- and right-hand panels).

Fig. 6b shows an example where we plot the AI-derived spectral density amplitudes $S_{\text{AI}}^{(i)}(\omega_0)$ at two select frequencies $\omega_0 = 2\pi \times 10$ and $2\pi \times 25$ kHz. The network manages to correctly reproduce the anticipated pattern in both instances, indicating a faithful derivation of the varying spectral densities throughout the data set. Note that although the synthetic noise field is predominantly colored within each diagonal block, the central frequencies are different, hence making it possible to selectively expose one or the other (a form of spectroscopic image contrast). These noise spectroscopy patterns portend experiments where a spin probe gathers compositional or dynamical information at different sites as it spectrally probes magnetic fluctuations across the sample. Even if a proof of principle, the above example highlights the potential of our approach, requiring only a comparatively small data set and hence only a fraction of the experimental time (consider implementing known high-accuracy methods of spectral density reconstruction at each site in the array).

## III. DISCUSSION

In summary, we leverage recent advances in machine learning and optically detected magnetic resonance of NV centers in diamond to introduce an AI-assisted approach to noise spectroscopy. In the present implementation, the sensing protocol takes the form of a Hahn-echo sequence, which simultaneously mitigates the undesired influence from nuclear spins in the host crystal while keeping NVs sensitive to magnetic field fluctuations within a broad, practically relevant frequency range. With the help of a convolutional auto-encoder network architecture, we derive spectral densities that reasonably approach those imprinted in the magnetic noise we engineer. We also find that the network performance is robust to the presence of experimental noise in the input observations, a behavior we associate with the network's ability to identify non-local features in the spin-probe response. Capitalizing on these findings, we carried out a proof-of-principle demonstration of spectroscopic imaging, a stepping stone towards integrating noise spectroscopy and precision imaging at the nanoscale. This approach to spectroscopic sensing should allow alternative forms of imaging contrast and thus images with different information content.

In this same vein, it is worth noting that the experiments in Fig. 6 — successively exposing the same group of NVs to traces of independently engineered noise — ignore the cross-spectra that can result from spatial correlations[37-39] (a situation akin to a scanning probe measurement). More generally, however, one can envision DNN-based strategies able to expose causal correlations between environmental dynamics at distinct locations or the occurrence of propagation of noisy signals. Such strategies could leverage devices relying on multiple probes that can be monitored in parallel, e.g., in the form of multi-tip diamond cantilevers[40], or homogeneous ensembles of shallow NVs illuminated by separate lasers[41]. Our work could also prove relevant to quantum information processing in its ability to quickly identify error models affecting the performance of spin qubits and, correspondingly, inform the search for effective dynamical decoupling protocols[42-45].

Since noise contributions at lower frequencies are typically dominant, we constrained our AI-assisted protocol to reconstructing spectral densities solely with the use of a Hahn-echo protocol (whose associated filter function is singularly broad). In this sense, the high predictive accuracy we attain is somewhat surprising as a broad frequency range tends to increase the degeneracy of the solutions for a given coherence decay. We hypothesize that the training dataset plays a fundamental role here: The network is trained to distinguish curves with $T_2$ times ranging from a 50 to 500 μs and engineered noises with frequency components in the order of units to hundreds of kHz. This combination of parameters spans a sufficiently large space, allowing the network to distinguish between nearly degenerate solutions and attain high precision.

While we limited our analysis to external noise sources, the AI-derived spectral density of the native noise — as obtained from the "Reference" NV response in Fig. 3b, not shown here for brevity — shows excellent agreement with that expected for a central spin model in a nuclear spin bath[46]. We warn, however, that a generalization to other, unknown environments is not straightforward: Since the noise spectrum producing a coherence decay for a given filter function is not uniquely determined, the network could, in principle, produce a noise spectrum that is significantly different from the target spectrum but that still reproduces the input coherence decay to a high degree of accuracy. Therefore, whilst the reproduction of the coherence decay allows a sense check of network performance, it does not absolutely guarantee accuracy. Bayesian neural network methods would prove a sensible extension to this work whereby a network would output a probability distribution over noise spectra rather than a single spectrum.

There is a natural limitation of the network when it comes to predicting frequencies outside the range given by the time vector of the coherence decay, which determines the probing frequencies of the filter function. Whilst the network has some ability to extrapolate outside these frequency limits, accuracy can be expected to fall. In the same vein, the network tends to predict results that match its training history, so when exploring unknown noise sources care must be taken to train the network on as diverse a set as is reasonably possible. For practical purposes, this initial work considered only a subset of all conceivable spectral densities, though we see no fundamental limitations to including more complex functional forms (e.g., multi-color noise). Along related lines, our work can be immediately extended to accommodate multi-pulse trains (i.e., $n \gg 1$), more suitable for probing higher-frequency components in $S(\omega)$, though at the expense of a narrower bandwidth.

An interesting extension would be the application of AI-methods to reconstruct spectral densities derived from spin-lattice relaxation measurements; note such work would be complementary because $T_1$ times are typically sensitive to processes occurring at a considerably higher frequency (roughly matching that defined by the spin transition of interest). We warn, however, that this class of experiments — even if adapted to the rotating frame via spin-locking techniques[47] — require control over different experimental parameters, and hence an alternative neural network architecture and training protocol would be in order.

Lastly, we mention that because spectral densities capture the underlying physical processes driving the fluctuations in the time-dependent noise field, a classification of $S(\omega)$ into different categories may suffice to inform the experimenter about differences between adjacent sections in a sample. This approach bodes well to the most widespread network architectures, conceived and optimized to separate diverse input data into different categories.

## IV. MATERIALS AND METHODS
### Experimental details

Magnetometry experiments are carried out with a small ensemble of NV centers in diamond. Formed by a substitutional nitrogen and a vacancy, the NV features a spin-1 ground-state triplet with a zero-field splitting $\Delta = 2.87$ GHz between the $m_S = 0$ and $\pm 1$ states. Optical spin initialization is possible at room temperature thanks to a spin-dependent excitation cycle that preferentially moves population from the $m_S = \pm 1$ to the $m_S = 0$ state upon intermediate shelving in a singlet manifold. The same mechanism also allows for optical readout of the ground spin state since shelving in the long-lived singlet reduces the photon emission rate[2].

The schematics in Fig. 1a lays out our experimental platform. We use a custom-made optical microscope with a green laser excitation (532 nm, 680 μW) to probe a diffraction-limited ensemble of NVs in a [111] diamond crystal with an estimated N (NV) concentration of 1 ppm (20 ppb). Spin manipulation in the presence of a bias field $B = 36$ mT normal to the sample — and thus parallel to one of the NV directions — is carried out via microwave (MW) pulses resonant with the $m_S = 0 \leftrightarrow m_S = -1$ transition; we use a thin (25 μm diameter) copper wire in contact with the diamond surface as the MW source.

Our experiments focused on bulk, not shallow, NVs out of convenience: The longer spin coherence lifetimes of bulk NVs allow us to make the effects of engineered noise dominant without resorting to impractically large input currents. Experimentally, this latter condition is not trivial to attain because the number of turns in the source coil (and hence its inductance) must be kept to a minimum so as to ensure a linear system response over a broad bandwidth.

### Noise Generation

Two types of noise were designed to evaluate the artificial-intelligence-based approach: colored noise and pink noise. The former one is built from an uncorrelated noise with a Gaussian distribution which is transformed by recursive methods[28] into a correlated noise that encodes a power spectrum $\propto 1/\omega^2$ with a bandwidth $\Delta\omega$. Next, the whole signal is multiplied by a cosine function with periodicity $v_0$, which leads to a colored noise $\propto 1/(\omega - \omega_0)^2$ and bandwidth $\Delta\omega$. On the other hand, the simpler pink noise ($\propto 1/\omega$) is generated with a built-in MATLAB function.

In order to deliver the noise signal to the loop, we use an AWG Tabor Electronics SE5082 with a sampling rate ranging from $5\times 10^7$ to $5\times 10^9$ samples/s and a memory of $32\times 10^6$ data points. Choosing a low sampling rate and using the full memory, the noise length is bounded between 0.5 and 1 s, which is shorter than the measurement time. In our experiments, we accumulate the NV signal in the presence of the same block of engineered noise until reaching a desired signal-to-noise ratio (SNR). We subsequently repeat the same protocol for several new noise blocks; the number of blocks is chosen so that, when analyzed as continuous time series, the spectral density $S_M(\omega)$ derived from the corresponding correlation function reasonably converges to the engineered target $S_C(\omega)$. While this strategy introduces undesired low-frequency components, it avoids the need to continuously feed the AWG with new synthetic noise, a technically demanding proposition. Note that these unwelcomed low-frequency components do not overlap with the region of interest (~1 kHz and up). As it can be seen in Fig. 1 (c), the resulting noise transmitted by the AWG (made out of one single noise block) already reproduces the desired shape and features of the designed power spectrum.

### DNN architecture

To implement our machine learning technique, we resort to the Keras application programming interface within the deep learning framework (Tensorflow 2) developed by Google. The networks used here build on those developed in Ref. [27] and are trained in a similar manner, namely, we use synthetic data, with coherence decay-noise spectrum pairs respectively functioning as input $X$ and target $Y$ data for the network. However, unlike the prior work — relying on a recurrent neural network (RNN) with a long-short term memory (LSTM) — here we use a convolutional auto-encoder.

We identify in our network two sections, which we refer to as the "encoder" and the "decoder". The encoder is formed by three convolutional layers based on rectified linear unit (ReLu) activations, each followed by a max pooling layer that reduces the dimensions of the input. The decoder comprises three convolutional layers also featuring ReLu activations followed by an up-sampling layer to expand the encoded data to the dimensions of the target. A final dense layer is used to match the output dimensions exactly using either linear or exponential activation. The former can assist in training time as the target noise spectra have values that often exceed $10^6$, greatly above the typical values between -5 and 5 that the network is initialized to output.

The network takes in an input vector of coherence values, with no explicit knowledge of their associated time vector. This time vector is implicit during training and so experimental data must use the same time vector. This limitation is mitigated by the use of interpolation (used extensively in this work) which can scale a coherence decay to the required time vector, provided that the experimental data has sufficient information in the relevant time range.

For training purposes, we generate approximately 80,000 spectral density curves, which we distribute across four different monotonic noise forms ($1/f$, Lorentzian, double Lorentzian, and $1/f$ + Lorentzian). Additional colored noise peaks are then added to these, with approximately 20% left as purely monotonic. The corresponding coherence decay curves are subsequently derived analytically using Eq. (1). These coherence decay curves then have Gaussian noise applied to them. This step is repeated six times with different noise to give approximately 500,000 total examples, 80% of which are used for training and 20% of which are reserved as a test set. The network is trained for 100 epochs, with each epoch representing an iteration through all training data. The training time is dependent on the hardware used, with an Nvidia K80 system requiring approximately 6 hours and an Nvidia V100 requiring < 1 hour. We use the one-cycle learning technique, whereby the learning rate is increased from a low initial value to a peak before being reduced again.

### V. ACKNOWLEDGMENTS.

We acknowledge discussions with Sara Mueller, Sara Ostadabbas, Amirreza Farnoosh, and Zhouping Wang at early stages of this work. F.M., D.P., and C.A.M. acknowledge support by the U.S. Department of Energy, Office of Science, National Quantum Information Science Research Centers, Co-design Center for Quantum Advantage (C2QA) under contract number DE-SC0012704; they also acknowledge access to the facilities and research infrastructure of the NSF CREST IDEALS, grant number NSF-HRD-1547830. This research was supported, in part, under National Science Foundation Grants CNS-0958379, CNS-0855217, ACI-1126113 and the City University of New York High Performance Computing Center at the College of Staten Island.